 \documentclass[apj]{emulateapj}

\shorttitle{Search for Hyper IR--Luminous DOGs selected with WISE and SDSS}
\shortauthors{Toba et al.}

\begin{document}


\title{Search for Hyper Infrared--Luminous Dust Obscured Galaxies \\ selected with WISE and SDSS}


\author{Y. Toba\altaffilmark{1} and T. Nagao\altaffilmark{1}} 





\altaffiltext{1}{Research Center for Space and Cosmic Evolution, Ehime University, Bunkyo-cho, Matsuyama, Ehime 790-8577, Japan}
\email{toba@cosmos.phys.sci.ehime-u.ac.jp}


\begin{abstract}
We aim to search for hyperliminous infrared (IR) galaxies (HyLIRGs) with IR luminosity $L_{{\rm IR}}$ $>$ 10$^{13}$ $L_{\sun}$ by applying the selection method of Dust Obscured Galaxies (DOGs).
They are spatially rare but could correspond to a maximum phase of cosmic star formation and/or active galactic nucleus (AGN) activity, hence they are a crucial population for understanding the star formation and mass assembly history of galaxies.
Combining the optical and IR catalogs obtained from Sloan Digital Sky Survey (SDSS) and Wide-field Infrared Survey Explorer ({\it WISE}), we performed the extensive HyLIRGs survey; we selected 5,311 IR-bright DOGs with $i$ -- [22] $>$ 7.0 and flux at 22 $\micron$ $>$ 3.8 mJy in 14,555 deg$^2$, where $i$ and [22] are $i$-band and 22 $\micron$ AB magnitudes, respectively.  
Among them, 67 DOGs have reliable spectroscopic redshifts that enable us to estimate their total IR luminosity based on the SED fitting.
Consequently, we successfully discovered 24 HyLIRGs among the 67 spectroscopically-confirmed DOGs. 
We found that (i) $i$ - [22] color of IR-bright DOGs correlates with the total IR luminosity and (ii) surface number density of HyLIRGs is $>$ 0.17 deg$^{-2}$.
A high fraction ($\sim$ 73\%) of IR-bright DOGs with $i$ - [22] $>$ 7.5 shows $L_{{\rm IR}}$ $>$ 10$^{13}$ $L_{\sun}$, and the DOGs criterion we adopted could be independently-effective against ``W1W2-dropout method'' based on four WISE bands, for searching hyper IR luminous populations of galaxies.
\end{abstract}





\section{Introduction}
Galaxies whose infrared (IR) luminosity exceeds 10$^{12}$ $L_{\sun}$ and 10$^{13}$ $L_{\sun}$ have been termed ultraluminous infrared galaxies \citep[ULIRGs:][]{Sanders} and hyperliminous infrared galaxies \citep[HyLIRGs:][]{Rowan-Robinson}.
Recently, galaxies whose IR luminosity exceeds 10$^{14}$ $L_{\sun}$ \citep[so-called extremely-luminous infrared galaxies, ELIRGs:][]{Tsai} have been discovered.
Their IR luminosity is thought to be generated by star formation (SF), active galactic nucleus (AGN) activity, or both.
Although ULIRGs/HyLIRGs are very rare populations in the local Universe, their relative abundance in galaxies is much higher at higher redshift. 
Accordingly they dominate the IR energy density at redshifts $z \sim$ 2, corresponding to the peak epoch of the cosmic SF and AGN activities \citep[e.g.,][]{Caputi,Magnelli,Goto}.
Therefore, it is important to search for IR luminous galaxies such as ULIRGs and HyLIRGs for understanding the galaxy formation and evolution and connection to their supermassive black holes (SMBHs).

One of the simple ways to select those IR luminous objects at $z \sim$ 2 is to use the optical and IR color, $R$ - [24] $>$ 7.5  where $R$ and [24] are $R$-band (0.65 $\micron$) and {\it Spitzer}/MIPS \citep{Werner,Rieke} 24 $\micron$ AB magnitude, respectively. 
This extreme optical--IR color indicates the presence of plenty of dust heated by active SF, AGN, or both, and the bulk of the optical and ultraviolet (UV) emission from them is absorbed by the surrounding dust.
Galaxies selected through this method is termed dust obscured galaxies \citep[DOGs:][]{Dey}.
Some surveys with the {\it Spitzer} space telescope have discovered DOGs and investigated their properties \citep[e.g.,][]{Houck,Brand,Brodwin,Desai,Melbourne_09,Bussmann_09,Bussmann_11,Bussmann_12,Melbourne}.
Recently, \cite{Toba} have discovered 48 DOGs whose IR luminosity is equivalent to that of HyLIRGs with the Hyper Suprime-Cam \citep[HSC:][]{Miyazaki} on the Subaru telescope and the Wide-field Infrared Survey Explorer \citep[{\it WISE}:][]{Wright}, and reported their statistical properties.
However, since the survey area of these studies is limited to $\sim$9 deg$^2$ at maximum, it is quite difficult to collect an enormous number of HyLIRGs whose volume densities are extremely low, although they are expected to be crucial population in terms of co-evolution of galaxies and SMBHs \citep[see ][]{Hopkins,Narayanan}.

In this paper, we present an effective method of searching for HyLIRGs by performing the largest DOGs survey based on {\it WISE} and Sloan Digital Sky Survey \citep[SDSS:][]{York}.
The large survey area of these catalogs is a key for identifying statistically significant samples of rare yet luminous sources.
Throughout this paper, the adopted cosmology is a flat universe with $H_0$ = 70 km s$^{-1}$ Mpc$^{-1}$, $\Omega_M$ = 0.3, and $\Omega_{\Lambda}$ = 0.7. Unless otherwise noted, all magnitudes refer on the AB system. 

\section{Data and analysis}
\subsection{Sample selection}
\begin{figure}
   \centering
   \includegraphics[width=0.45\textwidth]{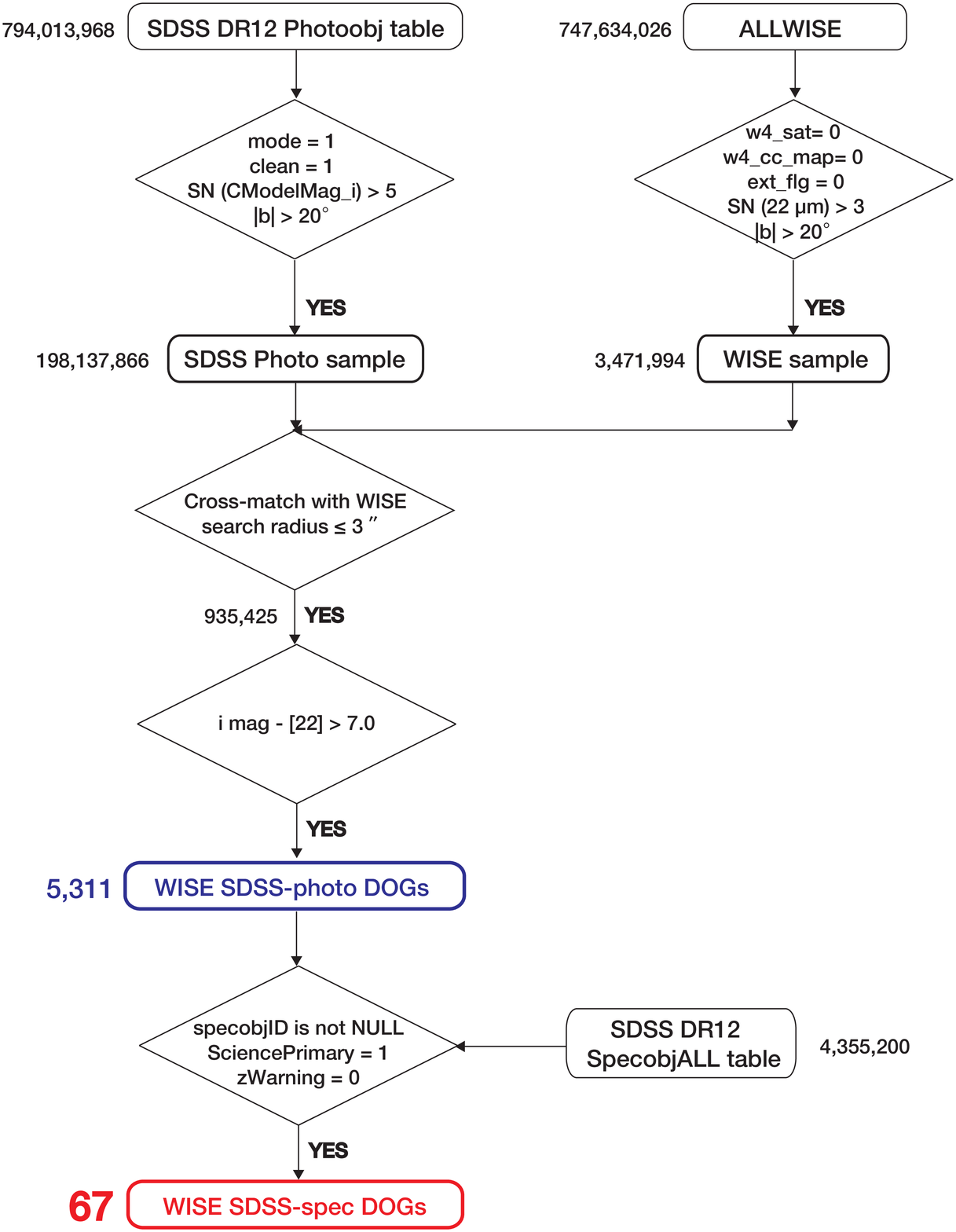}
   \caption{Flow chart of our DOGs selection process. Numbers in this figure denote the number of selected objects at each step.}
   \label{Sample}
   \end{figure}

We selected DOGs based on the {\it WISE} and SDSS catalogs.\footnote{For the selection process, we employed the TOPCAT based on the Starlink Tables Infrastructure Library (STIL), which is an interactive graphical viewer and editor for tabular data \citep{Taylor}.}
{\it WISE} performed an all-sky survey at 3.4 $\micron$ (W1), 4.6 $\micron$ (W2), 12 $\micron$ (W3), and 22 $\micron$ (W4) with angular resolutions of 6.1, 6.4, 6.5, and 12.0 arcsec, respectively \citep{Wright}.
We utilized a latest catalog, ``AllWISE'' Data Release \citep{Cutri} in which the 5$\sigma$ photometric sensitivity for 3.4, 4.6, 12, and 22 $\micron$ is better than 0.054, 0.071, 1, and 6 mJy
(corresponding to 19.6, 19.3, 16.4, and 14.5 mag), respectively.  
{\it WISE} catalog contains the Vega magnitude of each source, and we converted these to AB magnitude, using offset values $\Delta m$ ($m_{\mathrm{AB}}$ = $m_{\mathrm{Vega}}$ + $\Delta m$) for 3.4, 4.6, 12, and 22 $\mu$m of 2.699, 3.339, 5.174, and 6.620, respectively, according to the Explanatory Supplement to the AllWISE Data Release Products.\footnote{http://wise2.ipac.caltech.edu/docs/release/allwise/expsup/}
SDSS Data Release 12 (DR12) is the final data release of the SDSS-III, containing all SDSS observations  \citep{Alam}.
This catalog contains 469,053,874 unique and primary sources imaged over 14,555 deg$^2$ in five bands ($u$, $g$, $r$, $i$, and $z$) and 4,355,200 spectra.
The limiting magnitudes (95 \% completeness for point sources) are 22.0, 22.2, 22.2, 21.3, and 20.5 for $u$, $g$, $r$, $i$, and $z$, respectively.
Note that magnitudes in $u$- and $z$- band are slightly different to AB magnitude, we thus used $\Delta m$ = -0.04 and 0.02 for $u$- and $z$-band, respectively.\footnote{http://www.sdss.org/dr12/algorithms/fluxcal\#SDSStoAB}
The flow chart of our sample selection process is shown in Figure \ref{Sample}.
   
We first created the ``clean'' subsample for each catalog with considering the detection and photometry flags that ensure reliable detection and photometry.
To avoid the contamination from galactic stars, we narrowed our sample to sources at high galactic latitudes, $\vert$b$\vert$ $>$ 20$\arcdeg$.
In addition, we extracted sources with signal-to-noise ratio (SN) grater than 5 and 3 at $i$-band and 22 $\micron$, respectively, 
which yielded 198,137,866 SDSS sample and 3,471,994 {\it WISE} sample.
In this paper, we employed the profile-fit magnitude and CmodelMag for each source in the {\it WISE} and SDSS catalogs, respectively, which traces total flux (hereinafter we used $u$, $g$, $r$, $i$, and $z$ as a shorthand alias for CmodelMag).
We do not correct galactic extinction for SDSS photometry since our sample is located at relatively high galactic latitudes.
We then cross-identified {\it WISE} sources with SDSS sources with a search radius of 3$\arcsec$, and extracted 935,425 objects (hereinafter WISE--SDSS objects).
Note that 61,659 {\it WISE} objects ($\sim$2 \% of the matched sources) has more than one SDSS candidates of counterparts due to significant difference of the angular resolution between them.
We choose the nearest SDSS object as a counterpart for those cases.
For these 935,425 WISE--SDSS objects, we adopted the DOGs selection criterion: $i$ - [22] $>$ 7.0 \citep{Toba}\footnote{In that paper, we adopted a prior cult (i.e., $i$ - $K_{\rm s}$ $>$ 1.2) for HSC sources before cross-matching with WISE to avoid the false identification due to the significant difference of their angular resolutions. In the case of the SDSS and {\it WISE}, however, the probability of the false identification is much smaller because the detection limit of the SDSS data is much shallower than that of the HSC. We thus just cross-identified the SDSS and {\it WISE} without adopting any prior cut.}, where $i$ and [22] represent $i$-band (0.75 $\micron$) and 22 $\micron$ band magnitude, respectively.
As a result, 5,311 DOGs (hereinafter WISE--SDSS photo DOGs) were selected.
Note that the main difference of the WISE--SDSS photo DOGs and classical DOGs discovered by \cite{Dey} is the mid-IR (MIR) flux; we selected DOGs with SN (22 $\micron$) $>$ 3 corresponds to $\sim$ 3.8 mJy that is much brighter than 0.3 mJy at 24 $\micron$ selected by \cite{Dey}.
We focus on these ``IR-bright DOGs'' with $i$ -[22] $>$ 7.0 and flux at 22 $\micron$ $>$ 3.8 mJy.
For them, we extracted DOGs with reliable spectroscopic information with {\tt SciencePrimary} = 1 and {\tt zWarning} = 0.
Finally 67 DOGs (hereinafter WISE--SDSS spec DOGs) were selected through these steps.
    
   \begin{figure}
   \centering
   \includegraphics[width=0.45\textwidth]{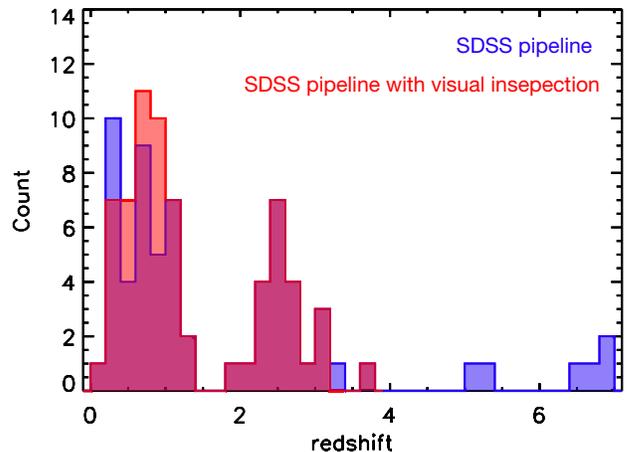} 
   \caption{Redshift distribution of 67 WISE--SDSS spec DOGs. The blue histogram represents redshift determined from the SDSS pipeline. The red histogram represents improved redshift determined by the SDSS pipeline and visual inspection.}
   \label{z_dist}
   \end{figure}

   \begin{figure*}
   \centering
   \includegraphics[width=0.9\textwidth]{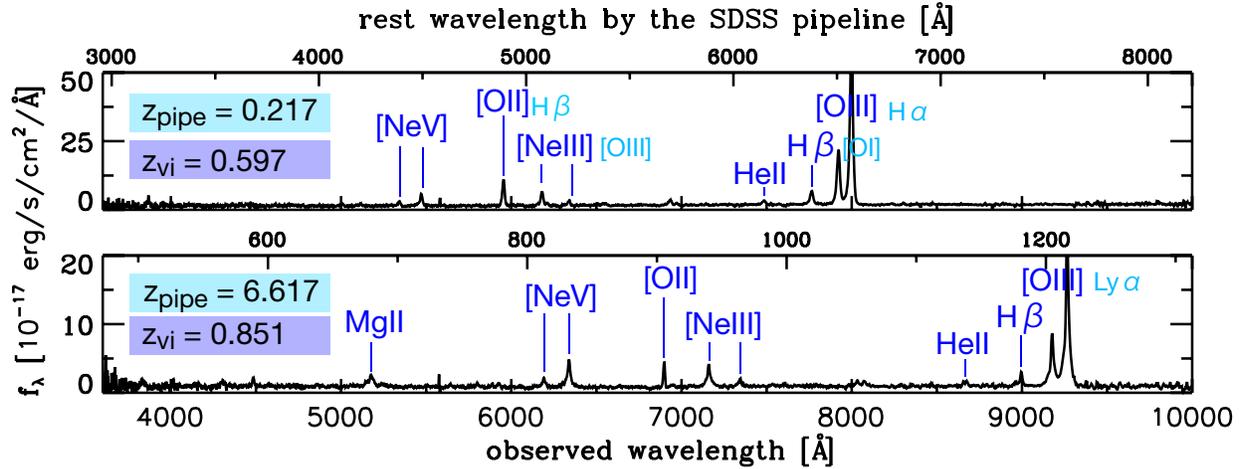}
   \caption{Examples of SDSS spectra with doubtful redshift. $z_{\rm pipe}$ means the redshift determined by the SDSS pipeline while $z_{\rm vi}$ means the redshift determined by visual inspection. The emission-line identifications based on redshifts determined by the SDSS pipeline and visual inspection are represented by cyan and blue colors, respectively.}
   \label{doubtful_z}
   \end{figure*}

Figure \ref{z_dist} shows the redshift distribution for WISE--SDSS spec DOGs.
Note that the pipeline of the SDSS sometimes misidentifies the redshift even for sources with {\tt zWarning} = 0.    
We hence checked each spectrum visually and re-identified the redshift for 10 objects that have doubtful redshift.
Figure \ref{doubtful_z} shows the example of spectra of objects whose redshift determined by the SDSS pipeline seems to be mis-identified.

\begin{figure}
	\centering
	\includegraphics[width=0.45\textwidth]{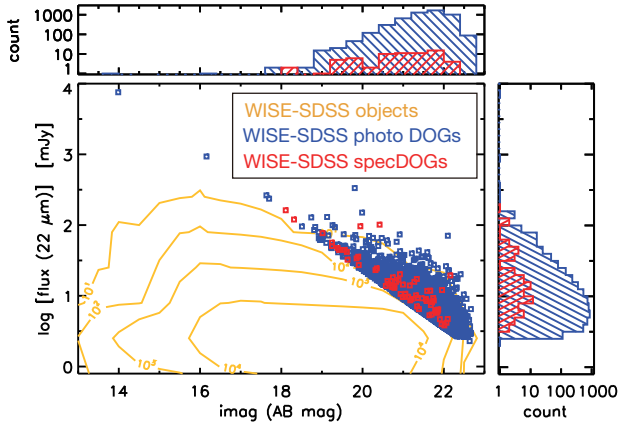} 	
	\caption{The $i$-band magnitude and 22 $\mu$m flux of our sample. The histograms of $i$-band magnitude and 22 $\mu$m flux are attached on the top and right, respectively. The yellow contours represent the number density of the WISE--SDSS objects (935,425 objects in total) in each 0.5 $\times$ 0.5 region on the $i$-band magnitude -- $\log$ [flux (22 $\mu$m)] plane. The blue and red points and histograms represent 5,311 WISE--SDSS photo DOGs and 67 WISE--SDSS spec DOGs, respectively.}
	\label{fhist}
\end{figure}

As shown in Figure \ref{z_dist}, the redshift distribution of the WISE--SDSS spec DOGs is bimodal with peaks around $z \sim$ 0.8 and $z \sim$ 2.6, which is in good agreement with that of \cite{Ross}.
Although they focused on the extremely red quasars selected from the {\it WISE} and SDSS, some of them should be overlapped with our sample. 
We will present the detailed spectroscopic properties of these WISE--SDSS spec DOGs in a forthcoming paper.  
Figure \ref{fhist} presents the 22 $\mu$m flux and $i$-band magnitude distributions for our sample. 
The average 22 $\mu$m flux densities for WISE--SDSS photo DOGs and spec DOGs are $\sim$ 13.3 and $\sim$ 23.0 mJy, respectively, while their average $i$-band magnitude are $\sim$ 21.5 and $\sim$ 20.9 mag, respectively.

\subsection{Cross-identification with AKARI}
\label{AKARI}
In order to obtain far-IR (FIR) data and perform a reliable SED fitting, we utilized data from {\it AKARI} satellite.
{\it AKARI} is the first Japanese space satellite dedicated to IR astronomy, that was launched in 2006 \citep{Murakami}.
{\it AKARI} performed an all-sky survey at 9, 18, 65, 90, 140, and 160 $\micron$ whose spatial resolution and sensitivity of AKARI are much better than those of the Infrared Astronomical Satellite \citep[{\it IRAS}:][]{Neugebauer,Beichman}, and MIR and FIR point source catalogs are publicly released.
The {\it AKARI}/IRC MIR all-sky survey catalog \citep{Onaka,Ishihara} contains 870,973 sources observed at 9 and 18 $\micron$ while {\it AKARI}/FIS FIR all-sky survey catalog \citep{Kawada,Yamamura} contains 427,701 sources observed at 65, 90, 140, and 160 $\micron$.
In particular, {\it AKARI} FIR survey provides the deepest data in terms of the FIR all-sky data, thus should be useful to derive the total IR luminosity.  
We cross-identified 67 WISE--SDSS spec DOGs with both catalogs where search radii for MIR and FIR catalogs are 6 and 60 arcsec, respectively.
As a result, however, our sample has no counterparts of {\it AKARI} MIR and FIR sources.
Nevertheless, this result can constrain the MIR and FIR fluxes for our sample as a upper limit.
In this study, we adopted 5$\sigma$ detection limit, 0.05, 0.12, 2.4, 0.55, 1.4 and 6.3 Jy, as an upper-limit at 9, 18, 65, 90, 140, and 160 $\micron$, respectively \citep{Ishihara,Kawada}.

\subsection{SED fitting}
\label{SED_fitting}
\begin{figure}
   \centering
   \includegraphics[width=0.45\textwidth]{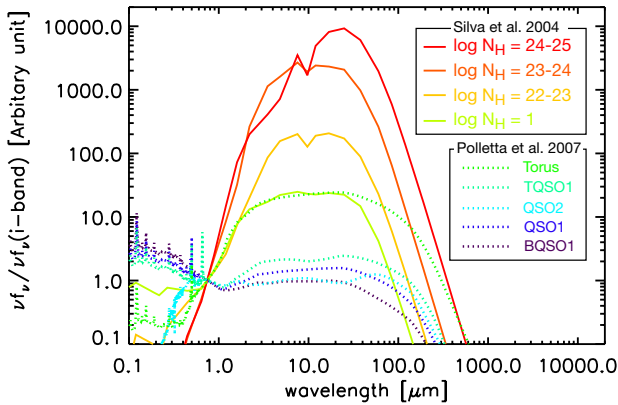}
   \caption{SEDs in $\nu f_{\nu}$ vs. $\lambda$ of the AGN templates we used in this study. Each SED is normalized at $i$-band (0.75 $\micron$). The solid lines are the SEDs presented by \cite{Silva} with $N_{\rm H}$ = 0 and 10$^{22-25}$ cm$^{-2}$ while the dotted lines are those presented by \cite{Polletta}}
   \label{Template}
   \end{figure}

We estimated the total IR luminosity, $L_{\mathrm{IR}}$ (8--1000 $\mu$m) for the 67 WISE--SDSS spec DOGs based on the SED fitting technique.
We employed the fitting code SEd Analysis using BAyesian Statistics ({\tt SEABASs}\footnote{http://xraygroup.astro.noa.gr/SEABASs/}: \citealt{Rovilos}) that provides up to three-component fitting (AGN, SF, and stellar component) based on the maximum likelihood method.
Since we do not have deep rest-frame MIR/FIR photometry responsible for SF activity (see Section \ref{AKARI}), we simply performed AGN and stellar component fitting for the 15 photometric points ($u$, $g$, $r$, $i$, $z$, and 3.4, 4.6, 9, 12, 18, 22, 65, 90, 140, and 160 $\micron$) with SDSS, {\it WISE}, and {\it AKARI} data, and estimated the total IR luminosity.
For the AGN templates, we utilized the library of \cite{Silva}, which contains torus templates with varying extinction ranging from $N_{\rm H}$ = 0 to $N_{\rm H}$ = 10$^{25}$ cm$^{-2}$.
We also used the library of \cite{Polletta} representing optically selected QSOs with different
values of IR/optical flux ratios (QSO1, TQSO1, and BQSO1) and two type 2 QSOs (QSO2, Torus) (see \citealt{Polletta} for more detail).
Figure \ref{Template} shows the SEDs of the AGN template we used.
For the stellar templates, {\tt SEABASs} gives a library of 1500 synthetic stellar templates from \cite{Bruzual} stellar population models with solar metallicity and a range of SF histories and ages, and each model are reddened using a \cite{Calzetti} dust extinction law.
Note that the uncertainties of the derived $L_{\rm IR}$ contains not only statistical error but also systematic error.
{\tt SEABASs} can calculate $L_{\rm IR}$ for ``every'' trial fit and estimate the likelihood value (corresponding to the chi-square) for each case, and provides us the uncertainties as the 2 $\sigma$ confidence interval.
Therefore the influence of the difference between the inputed SED templates on the derived $L_{\rm IR}$ is included in the uncertainty.
However, the above uncertainty does not consider the influence of the absence of the SF component when executing the SED fitting. We discuss this influence on the derived $L_{\rm IR}$ in Section \ref{LIR_err}.

\begin{figure*}
   \centering
   \includegraphics[width=0.9\textwidth]{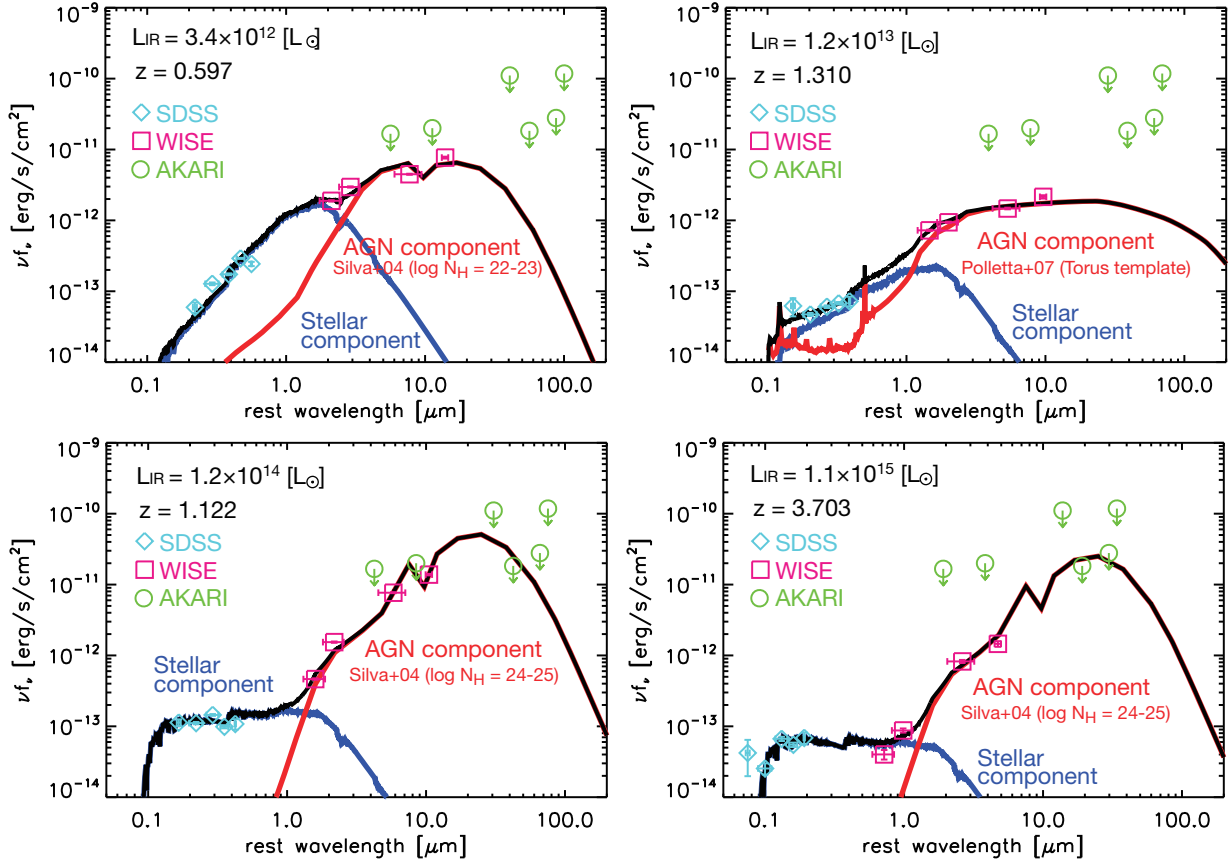}
   \caption{Examples of the SED fitting for our sample. The blue diamond and pink square represent the data from the SDSS and WISE, respectively. The green circle represents 5$\sigma$ upper limit obtained from AKARI data. The contribution from the stellar and AGN components to the total SEDs are shown in blue and red lines, respectively. The black solid line represents the resultant (the combination of the stellar and AGN) SEDs.}
   \label{SEABAS}
   \end{figure*}

Figure \ref{SEABAS} shows the example of the SED fitting.
The typical uncertainties of $L_{\rm IR }$ is 5-8 \% (but see Section \ref{LIR_err}).
The best-fit AGN template tends to favor the ``torus'' template presented by \cite{Silva} or \cite{Polletta}, which is consistent with the report by \cite{Tsai} based on the WISE-selected IR luminous sources.

\section{Result}
\begin{figure}
   \centering
   \includegraphics[width=0.45\textwidth]{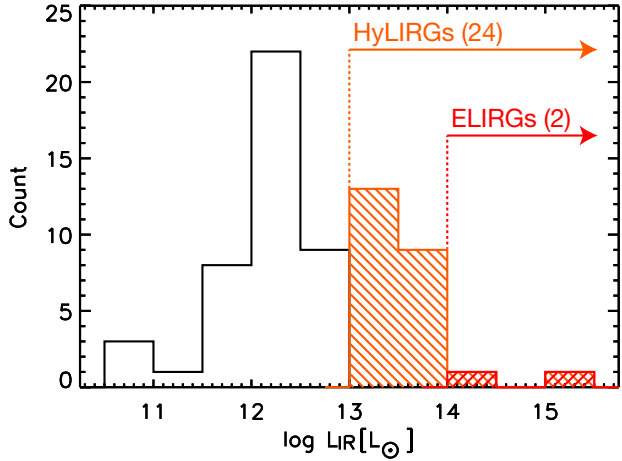}
   \caption{The distribution of the total IR luminosity for WISE--SDSS DOGs. The orange histogram represents $L_{\rm IR}$ for HyLIRGs while the red histogram represents that for ELIRGs. Numbers in parentheses denote the number of objects.}
   \label{LIR}	
   \end{figure}

\begin{figure*}
   \centering
   \includegraphics[width=0.9\textwidth]{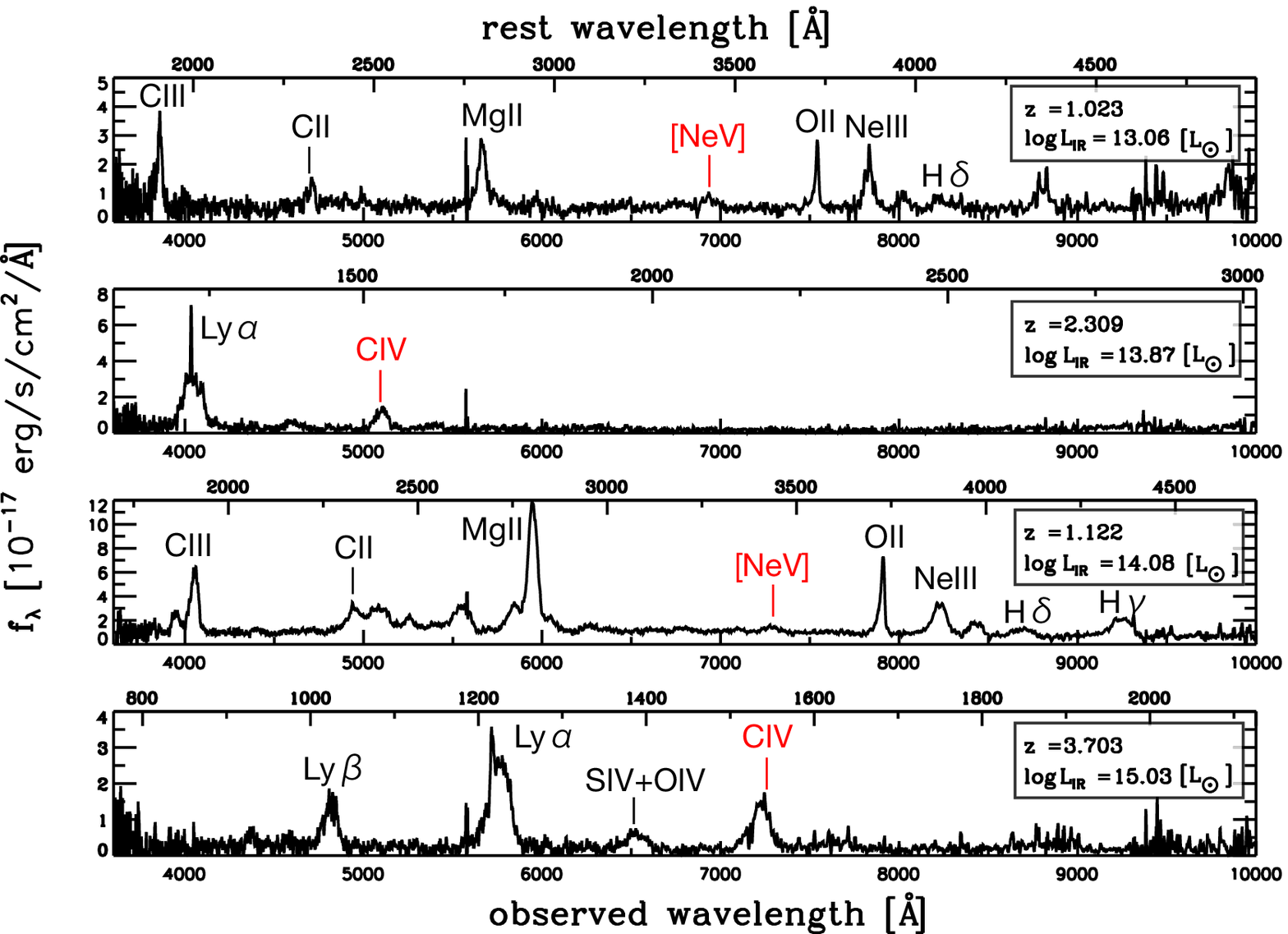}
   \caption{The SDSS spectra for 4 HyLIRGs/ELIRGs in our WISE--SDSS spec DOGs sample. Labels denoted by red color are indicators to identify AGNs (see the main text in detail).}
   \label{HyLIRGs}
   \end{figure*}

Figure \ref{LIR} shows the histogram of resultant total IR luminosity ($L_{\mathrm{IR}}$) for 67 WISE--SDSS DOGs.
Consequently, we successfully discovered 24 HyLIRGs.
Among those ``hyper-luminous DOGs'', 19 objects are $z > 2$ while 5 objects are $z < 2$.
In addition, 2 objects with $L_{\mathrm{IR}}$ = 1.2$^{+0.03}_{-0.04}$ $\times$ 10$^{14}$ L$_{\odot}$ and $L_{\mathrm{IR}}$ = 1.1$^{+0.10}_{-0.12}$ $\times$ 10$^{15}$ L$_{\odot}$ were identified to be ELIRGs (but see Section \ref{LIR_err}).
The examples of the SDSS spectra are shown in Figure \ref{HyLIRGs}.
Those 24 hyper-luminous DOGs are expected to harbor AGNs since we confirmed the presence of C~{\sc iv}$\lambda$1549 line for all of $z > $ 2 HyLIRGs and [Ne~{\sc v}]$\lambda$3426 line for all of $z <$ 2 HyLIRGs.
Since the ionization potentials of C~{\sc iv} and [Ne~{\sc v}] are 64.5 and 97.1 eV respectively, those lines cannot be present without AGNs.
Therefore, the AGN activity contributes more or less to the total IR luminosity.

\begin{figure}
   \centering
   \includegraphics[width=0.45\textwidth]{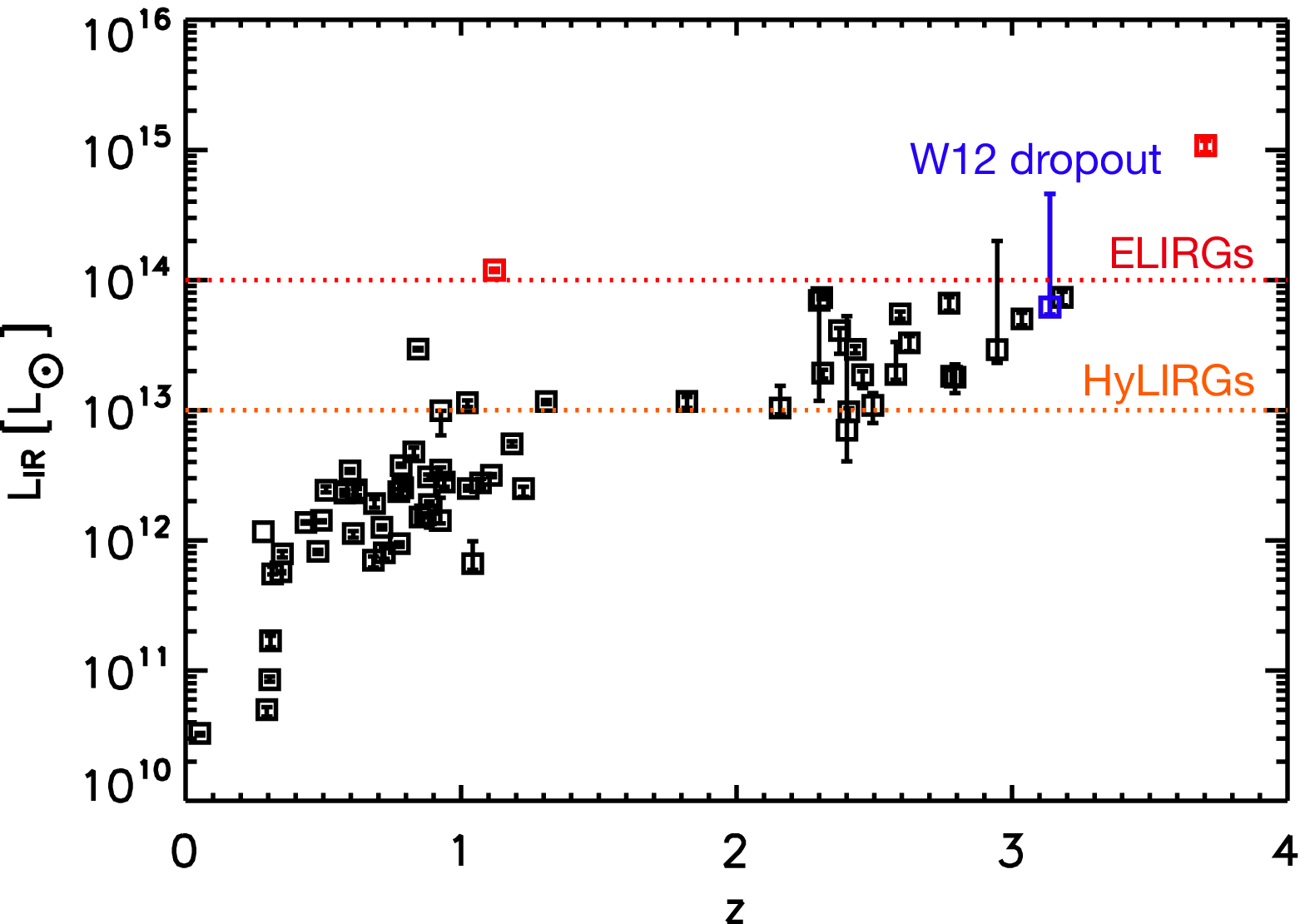}
   \caption{Total IR luminosities as a function of redshift. The error in $L_{\mathrm{IR}}$ includes the uncertainties of the SED fitting. The blue square represents a W12--dropout object (see Section \ref{HotDOGs}).}
   \label{z_L}
   \end{figure}

Figure \ref{z_L} shows the relation between the redshift and $L_{\mathrm{IR}}$ for WISE--SDSS spec DOGs.
Note that those IR luminosity could be a lower limit since we do not use the SF template when performing the SED fitting, although one should keep in mind there could be a large uncertainty due to the lack of good constraints at FIR (see Section \ref{LIR_err}).

\section{Discussion}

\subsection{Uncertainties in the derived infrared luminosity}
\label{LIR_err}
In this study, we derived $L_{\rm IR}$ based on the SED fitting given the SED template of stellar and AGN component, without considering the SF component.
However, one caution is that the current data do not always constrain the peak of the SED.
In the case of high-z objects, in particular, the peak wavelength of the SED contributed from AGNs is  beyond the longest detection at 22 $\micron$ (see, for example, the lower-right panel in Figure \ref{SEABAS}), which induces the large uncertainties of $L_{\rm IR}$.
If the SF component affects their 12 and 22 $\micron$ flux, the derived  $L_{\rm IR}$ has a potential not only of underestimate as described in Section \ref{SED_fitting}, but also of overestimate due to the lack of the data covering the peak of the SED of the AGN component.  
We discuss here how the AGN template will be affected and how the overall SED and derived $L_{\rm IR}$ will change correspondingly, when executing the SED fitting with or without SF component for our data.

We evaluate this effect based on the 113 DOGs presented by \cite{Melbourne} who also derived the $L_{\rm IR}$ for DOGs by performing the SED fitting.
The DOGs sample used for the SED fitting were originally discovered by deep optical and MIR imaging taken with NOAO Deep Wide-Field Survey \citep{Jannuzi} and {\it Spitzer}.
They have 11 photometries ($B_{\rm w}$, $R$, $I$, and 3.6, 4.5, 5.8, 8.0, 24, 250, 350, 500 $\micron$) where the optical, MIR, and FIR data are taken with KPNO, {\it Spitzer}/IRAC \citep{Fazio} and MIPS, and {\it Herschel}/SPIRE \citep{Pilbratt,Griffin} respectively, and have spectroscopic redshift. 
In particular, deep FIR data are used for determining their precise $L_{\rm IR}$, and photometric data and resultant $L_{\rm IR}$ for each DOG are available in \cite{Melbourne}.

First, we check the consistency of the $L_{\rm IR}$ derived from our method (i.e., the SED fitting with {\tt SEABASs}) and those in \cite{Melbourne}; we calculated the $L_{\rm IR}$ for the same data as \cite{Melbourne} by executing {\tt SEABASs} with stellar/AGN/SF components, and compared them with the $L_{\rm IR}$ given in \cite{Melbourne}.
For the SF components, we utilized a SED library presented by \cite{Polletta}.
   \begin{figure}
   \centering
   \includegraphics[width=0.45\textwidth]{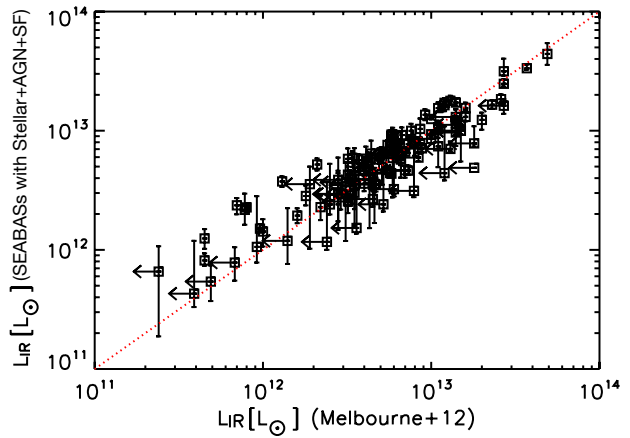}
   \caption{The comparison of $L_{\mathrm{IR}}$ derived by {\tt SEABASs} with Stellar/AGN/SF components and \cite{Melbourne}. The red dotted line is the one-to-one line.}
   \label{LIR_comp}
   \end{figure}
Figure \ref{LIR_comp} shows the comparison of $L_{\mathrm{IR}}$ derived by {\tt SEABASs} and \cite{Melbourne}, which indicates that they are roughly consistent with each other.

\begin{figure}
   \centering
   \includegraphics[width=0.45\textwidth]{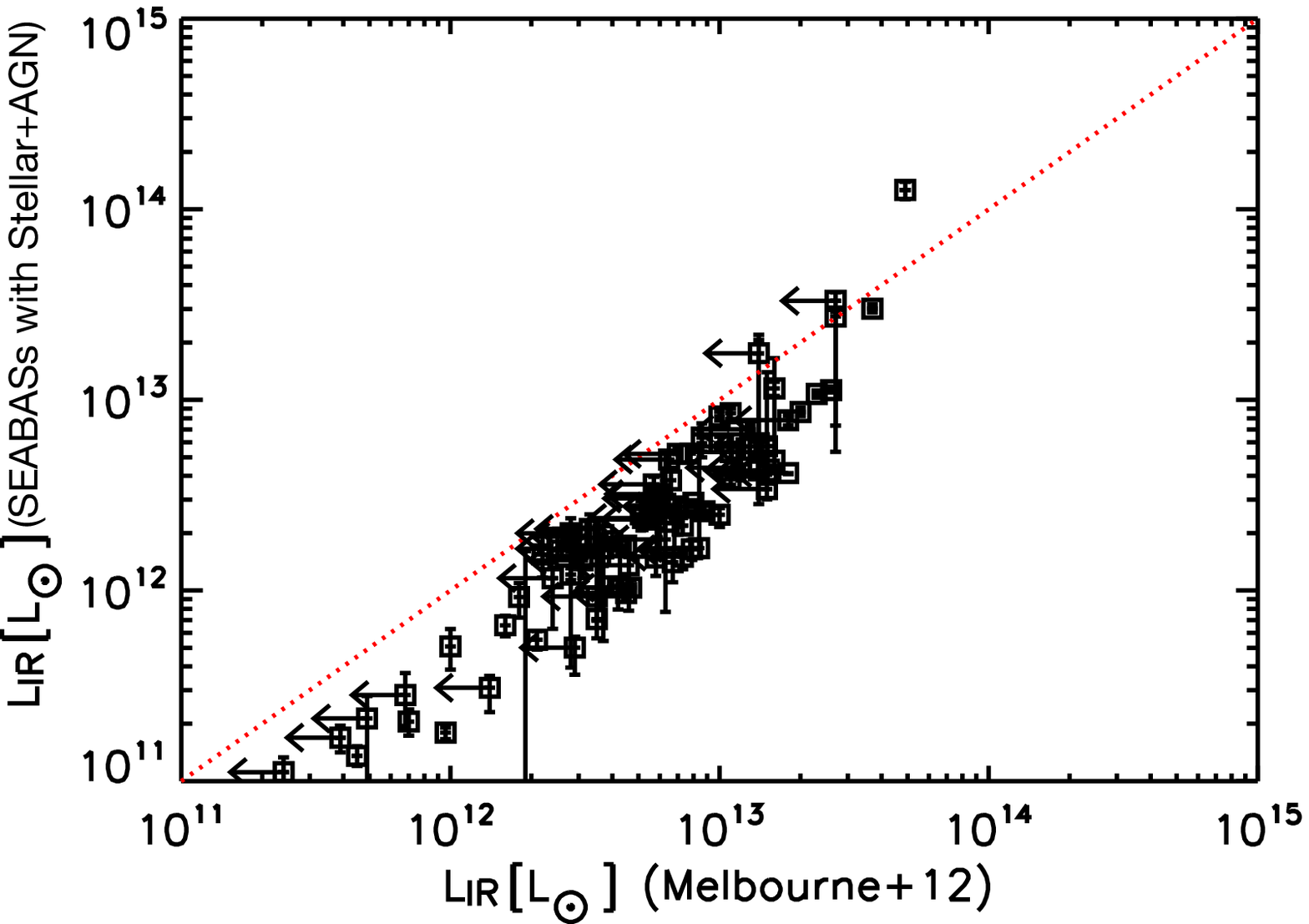}
\caption{The comparison of $L_{\mathrm{IR}}$ derived by {\tt SEABASs} without SF component and \cite{Melbourne}. The red dotted line is the one-to-one line.}
   \label{LIR_comp_AGN}
   \end{figure}

\begin{figure*}
   \centering
   \includegraphics[width=0.9\textwidth]{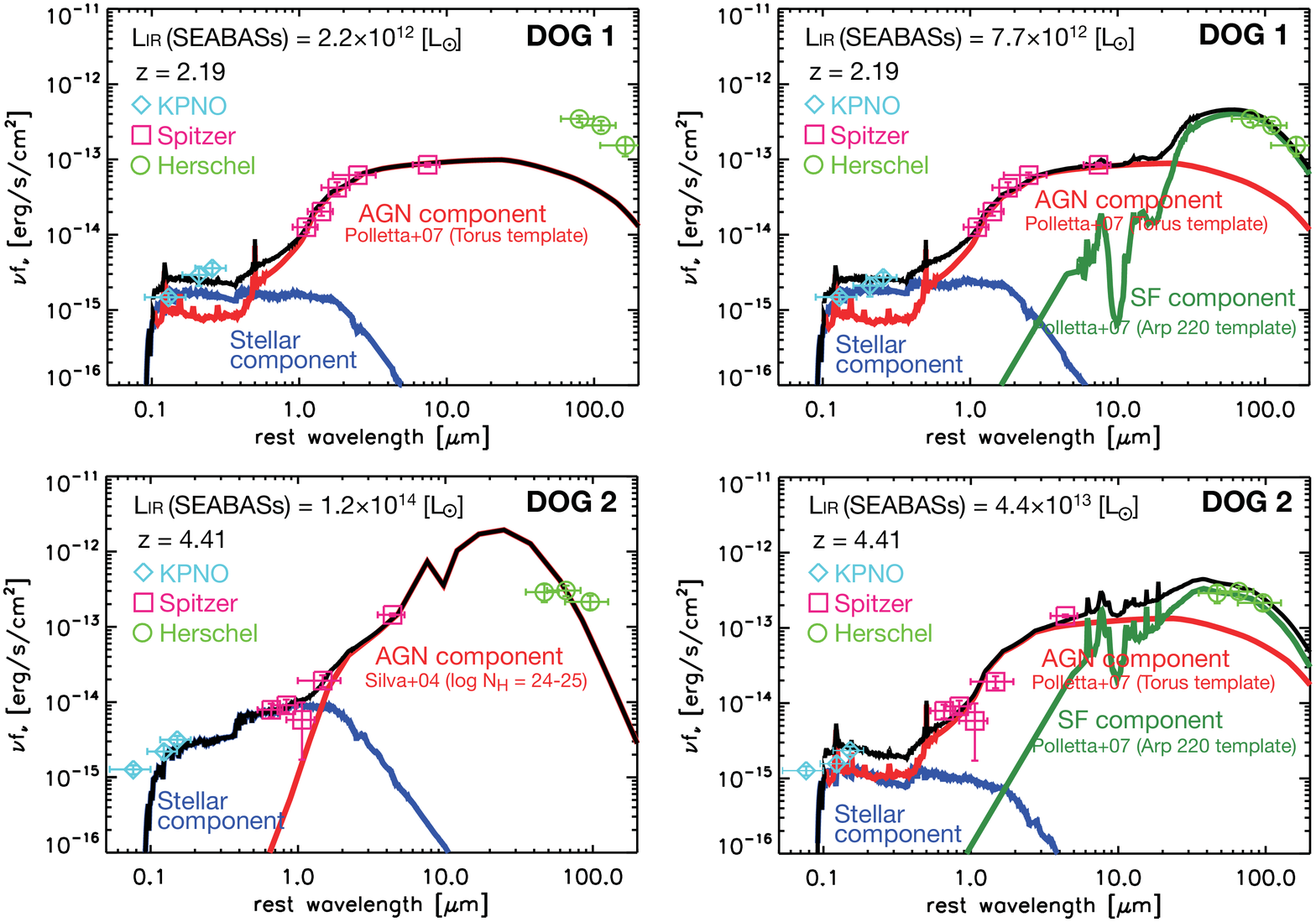}
   \caption{Examples of the SED fitting for 2 DOGs in \cite{Melbourne}. The upper and lower panels show an example of the underestimate and overestimate of the $L_{\mathrm{IR}}$, respectively. The blue diamond, pink square, and green circle represent the data from the KPNO, {\it Spitzer}, and {\it Herschel}, respectively. The contribution from the stellar, AGN, and SF components to the total SEDs are shown in blue, red, and green lines, respectively. The black solid line represents the resultant (the combination of the stellar and AGN) SEDs.}
   \label{LIR_comp_AGN_SB}
   \end{figure*}

Next, we calculate the $L_{\mathrm{IR}}$ based on {\tt SEABASs} without SF template (i.e., Stellar + AGN templates) for the same data and compare them with those in \cite{Melbourne}, as shown in Figure \ref{LIR_comp_AGN}.
This result indicates that the lack of SF component basically induces the underestimate of $L_{\mathrm{IR}}$ when performing the SED fitting, whereas the absence of the SF component has a potential of overestimate for DOGs with a very large $L_{\mathrm{IR}}$.
Figure \ref{LIR_comp_AGN_SB} shows an example of the SED fitting with or without SF component for 2 DOGs (we call them ``DOG 1'' and ``DOG 2'' tentatively).
The $L_{\mathrm{IR}}$ derived by \cite{Melbourne} is $7.3 \times 10^{12} L_{\odot}$ and $4.9 \times 10^{13} L_{\odot}$ for DOG 1 and DOG 2, respectively.
The $L_{\mathrm{IR}}$ derived by {\tt SEABASs} with SF component is $7.72 \times 10^{12} L_{\odot}$ and $4.42 \times 10^{13} L_{\odot}$ for DOG 1 and DOG 2, respectively, which is in good agreement with those of 
\cite{Melbourne}.
On the other hand, the $L_{\mathrm{IR}}$ derived by {\tt SEABASs} without SF component is $2.20 \times 10^{12} L_{\odot}$ and $1.27 \times 10^{14} L_{\odot}$ for DOG 1 and DOG 2, respectively, which means that we underestimate the IR luminosity for DOG 1 while we overestimate it for DOG 2. 
In this case, the absence of the SF component influence on the derived $L_{\mathrm{IR}}$ by a factor of 3-4.
In any case, one should keep in mind this uncertainty when discussing the $L_{\mathrm{IR}}$.

As discussed above, in the case of having the deep {\it Herschel} data, the derived $L_{\mathrm{IR}}$ is reliable with small uncertainties.
However, in this study, we performed the SED fitting for including {\it AKARI} data which give the upper limit of FIR flux.   
If we do not have deep {\it Herschel} data but have only much shallower {\it AKARI}/FIS data as this work, can AKARI data do a similar job to have good constraints to the IR luminosity?
To address this issue, we estimate the $L_{\mathrm{IR}}$ for KPNO + {\it Spitzer} + {\it AKARI}/FIR data for DOG 1 and DOG 2, and compare with those derived from KPNO + {\it Spitzer} + {\it Herschel}.
The resultant $L_{\mathrm{IR}}$ are $2.39 \times 10^{12} L_{\odot}$ and $2.96 \times 10^{14} L_{\odot}$ which is 0.31 and 6.70 times larger than those derived with {\it Herschel} data, respectively.
Therefore, the upper limit of {\it AKARI}/FIS photometry could not work very well for high $L_{\mathrm{IR}}$ objects because they are usually those with a fitted AGN peak at much longer wavelengths than 22 $\mu$m.  
In addition to the absence of SF component, one should keep in mind that the absence of deep FIR data also influences on the derived $L_{\mathrm{IR}}$.

\subsection{A new effective method for searching for HyLIRGs}
\label{discuss}
\begin{figure}
   \centering
   \includegraphics[width=0.45\textwidth]{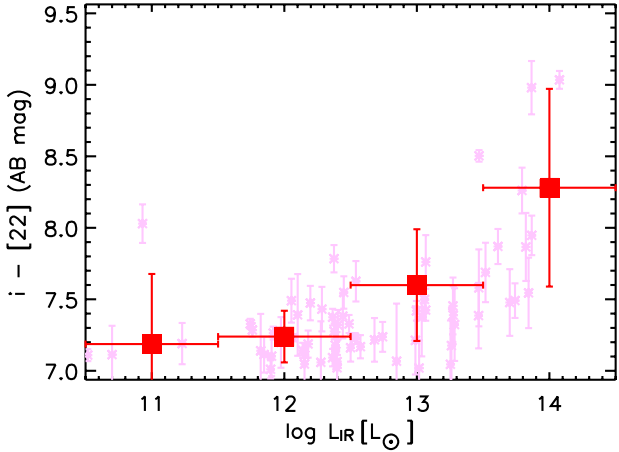}
   \caption{The relation between $L_{\mathrm{IR}}$ and $i$ - [22] color for WISE--SDSS spec DOGs. The pink asterisk represents the individual data while the red square represents the average and its standard deviation per log $L_{\rm IR}$ bin.}
   \label{i_22_LIR}
   \end{figure}

The relation between $i$ - [22] and $L_{\mathrm{IR}}$ as shown in Figure \ref{i_22_LIR} gives us a clue of effective search for HyLIRGs.
This figure indicates that objects with redder $i$ - [22] color tend to have higher infrared luminosity, although there is large scatter. 
In order to investigate how reliable this correlation quantitatively, we performed the Spearman's rank order test for individual data.
The resultant Spearman's rank coefficient for this correlation is 0.51, corresponding to a null hypothesis probability of 9.9 $\times$ 10$^{-6}$.
This means that $i$ - [22] and $L_{\rm IR}$ for IR-bright DOGs have a positive correlation with a high significance.
This correlation can be interpreted qualitatively as follows: 
The IR luminosity is generated by combination of SF, AGN, or both, and the contribution of AGNs to the IR luminosity increases with increasing IR luminosity \citep[e.g.,][]{Sanders,Yuan,Ichikawa}.
On the other hand, $i$ - [22] value mostly corresponds to the extinction, that correlates with SFR at larger stellar mass \citep[e.g.,][]{Zahid}.
In the scenario that ULIRGs/HyLIRGs are emerged through the major merger \citep[e.g.,][]{Kartaltepe}, the absolute contribution both of AGN and SF can be increased and HyLIRGs could correspond maximum phase in terms of AGN and SF activity \citep[see also ][]{Hatziminaoglou}.
This leads to a positive correlation between $L_{\mathrm{IR}}$ and $i$ - [22] color.
    \begin{figure}
   \centering
   \includegraphics[width=0.45\textwidth]{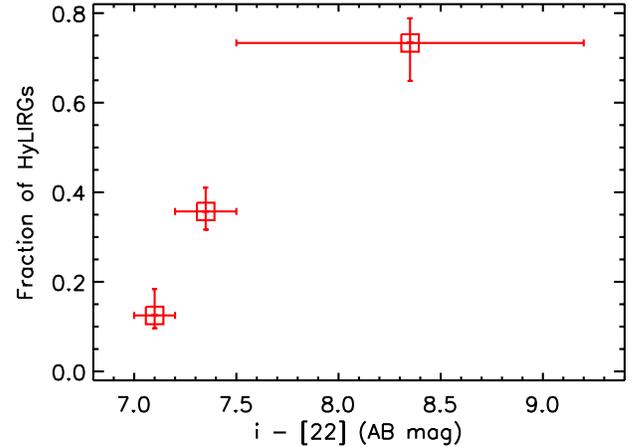}
   \caption{The fraction of HyLIRGs as a function of $i$ - [22] color for WISE--SDSS spec DOGs. Error in the fraction was estimated using binomial statistics (see \citealt{Gehrels}). The color interval is optimized to keep the data points more or less equal for each bin.}
   \label{i_22_Hyfrac}
   \end{figure}
   
Figure \ref{i_22_Hyfrac} shows the fraction of HyLIRGs as a function of $i$ - [22] color.
When focusing DOGs with $i$ - [22] $>$ 7.5, almost 70\% of objects are HyLIRGs, which is a valuable guideline for performing systematic surveys of HyLIRGs/ELIRGs.
The number of the SDSS--WISE photo DOGs with $i$ - [22] $>$ 7.5 is 1,327, thus this sample probably contains $\sim$ 930 HyLIRGs.
Note that the above correlation is seen in DOGs with 22 $\micron$ flux grater than 3.8 mJy, and thus our result may not be applicable for IR-faint DOGs.
Therefore further studies on the relation between the $i$ - [22] color and the HyLIRG fraction are needed to generalize this method for HyLIRG searches, that is beyond the scope of this paper.

Note also that IR luminosities of those ELIRGs are potentially amplified by effect of beaming and/or gravitational lensing even if the derived $L_{\rm IR}$ is reliable.
It is quite difficult to reject these possibilities quantitatively based on the currently available data.
So, we just mention here that some observational supports for possible HyLIRGs.
For the potential of the beaming effect, we checked their variability flag at 22 $\micron$ in ALLWISE catalog because beaming effect could induce the rapid variability as discussed in \cite{Tsai}.
As a result, 10 objects including possible two ELIRGs have {\it var\_flg} = 0, which means that they are expected to show no variability while 4 objects have 1 $<$ {\it var\_flg} $<$ 2, meaning that they could show very weak variability.
The remaining 10 objects have {\it var\_flg} = ``n'' that indicates insufficient or inadequate data to make a determination of possible variability.
Note that although the estimate of the flux variability is unreliable for extended source with {\it ext\_flg} $>$ 0 or for contaminated by image artifact with {\it cc\_flag} $\neq$ 0 in the pipeline of ALLWISE, our sample have {\it ext\_flag} = 0 and {\it cc\_flag} = 0, hence the estimate of the flux variability should be reliable.
For the potential of lensing effect, 
$i$ - [22] color dependence of IR luminosity indicates the extreme IR luminosities of those HyLIRGs are unlikely due to the lensing.
If their IR luminosities are just amplified by lensing, their $i$ - [22] color are expected to be comparable to those of ULIRGs, but indeed their $i$ - [22] color get redder.
Therefore, the influence of the beaming and lensing on $L_{\rm IR}$ for most HyLIRGs discovered in this work could be expected to be small.

\subsection{Relation with ``W1W2--dropouts''}
\label{HotDOGs}
Recently, some authors have reported that a method based on the {\it WISE} color is also useful to discover HyLIRGs and ELIRGs \citep[e.g.,][]{Eisenhardt,Wu,Tsai}.
Most of objects that are faint or undetected by {\it WISE} at 3.4 $\mu$m (W1) and 4.6 $\mu$m (W2) but are well detected at 12 $\mu$m (W3) or 22 $\mu$m (W4) can be classified as HyLIRGs, and some of them are ELIRGs.
These objects are termed ``W1W2--dropouts'' or ``Hot DOGs'' \citep{Eisenhardt,Wu}.
Since our sample of HyLIRGs/ELIRGs are selected based on {\it WISE} data, we checked whether our sample satisfy the criteria for W1W2--dropouts.

The most critical selection criteria for W1W2--dropouts are W1 $>$ 17.4 and either (i) W4 $<$ 7.7 and W2 -- W4 $>$8.2 or (ii) W3 $<$ 10.6 and W2 -- W3 $>$ 5.3 where all magnitudes are Vega magnitudes \citep[for details, see ][]{Eisenhardt}.
When adopting the above criteria to WISE--SDSS spec DOGs, only one object satisfied the criteria.
This object is located at $z$ = 3.14 and classified as HyLIRGs in our sample, which represents blue square in Figure \ref{z_L}.
The fact that almost all of our sample do not satisfy the criteria of W1W2--dropouts is reasonable because W1W2--dropouts are expected to be very faint in the optical due to their steep continuum, and thus relatively shallow imaging in the SDSS spectroscopic catalog could miss large amount of W1W2--dropouts.
In other words, most HyLIRGs selected based on $i$ - [22] color do not duplicate of those in W1W2--dropout method, suggesting that our selection method could be independently useful for selecting HyLIRGs.
We emphasize that those selection methods are complemental.
The W1W2--dropout method could be sensitive to heavily absorbed HyLIRGs while our method is sensitive to less or moderately aborted (compared with W1W2--dropouts) HyLIRGs.
In that sense, our method is complemental compared with the W1W2--dropout method.

\subsection{Surface number density of hyper-luminous DOGs}
We here estimate the surface number density of hyper-luminous DOGs (IR-bright DOGs that satisfy HyLIRGs' criterion in a strict sense) and compared it with that of other sample presented by \cite{Bridge} who selected HyLIRGs based on a {\it WISE} color similar to W1W2--dropouts with optical magnitude cut ($r \sim$ 22).
They reported that these objects have a surface density of $\sim$ 0.1 deg$^2$.
It should be noted that since DOGs are generally faint in the optical, our HyLIRGs search with SDSS could induce a significant sample incompleteness.
In addition, the selection function of the SDSS spectroscopic survey is complicated. 
In this paper, we hence simply estimate the lower limit of surface number density by assuming that
(i) the 22 $\mu$m flux of the hyper-luminous DOGs almost reaches the 100 \% completeness (i.e, completeness correction in terms of the {\it WISE} survey can be neglected) and (ii) the WISE--SDSS photo DOGs are approximately completely selected by at least the depth of the SDSS spectroscopic survey which is shallower than the photometric survey. 

Under these assumptions, the surface number density of hyper-luminous DOGs ($N_{\mathrm{HyLIRGs}}$) can be derived as the sum of the expected number of possible HyLIRGs in the WISE--SDSS photo DOGs sample and HyLIRGs discovered in this study, per each apparent $i$-band magnitude bin;
\begin{equation}
N_{\mathrm{HyLIRGs}} = \frac{1}{A} \sum_j^N S_j(m_i) \times n^P_j(m_i) + n^E_j (m_i) \;,
\end{equation}
where $N$ is the number of apparent $i$-band magnitude bin, $A$ is the survey area (14,555 deg$^2$) in this study and $S_j (m_i)$ is the scaling factor that is the ratio of the number of the hyper-luminous DOGs and the WISE--SDSS spec DOG each $j$ bin.
$n^P_j (m_i)$ and $n^E_j (m_i)$ are the number of the WISE--SDSS photo DOGs without spectroscopic information and hyper-luminous DOGs discovered in this study, in each $j$ bin, respectively.
As a result, we obtained the $N_{\mathrm{HyLIRGs}}$ $>$ 0.17 deg$^{-2}$, which is roughly consistent with those derived by \cite{Bridge}.

\section{Summary}
In this work, we performed to search for HyLIRGs with $L_{\rm IR} >$ 10$^{13}$ $L_{\sun}$ that are most likely to be traced the maximum phase in terms of the SF and AGN activity and thus important population for understanding the co-evolution of SMBH-galaxy.
We utilized the latest {\it WISE} and SDSS catalogs and selected 67 candidates of hyper-luminous DOGs with 22 $\micron$ flux $>$ 3.8 mJy based on the criterion of the DOGs selection ($i$ - [22] $>$ 7.0).
We then executed the SED fitting for the data obtained from {\it AKARI}, {\it WISE}, and SDSS to estimate their $L_{\rm IR}$.
The main results are as follows:
\begin{enumerate}
\item We succeed in discovering the 24 HyLIRGs including two possible ELIRGs with $L_{\rm IR} >$ 10$^{14}$ $L_{\sun}$.
\item Their $i$ - [22] color correlates with $L_{\rm IR}$; the detection rate of HyLIRGs is about 73 \% when extracting the IR-bright DOGs with $i$ - [22] $>$ 7.5, which is an useful method to search for these IR luminous galaxies.
\item Among 24 HyLIRGs, only one object satisfy the criteria of W12--dropouts whose $L_{\rm IR}$ are equivalent to those of HyLIRGs selected based only from WISE data, which means that the suggested method in this study could be independently-effective against W12--dropout method.
\item The surface number density of 24 hyper-luminous DOGs is $\sim$ 0.17 deg$^2$ that is rough consistent with that of previous works.
\end{enumerate}
This study is based on the WISE and SDSS spectroscopic data that photometric sensitivities are shallower than those of the SDSS photometric data, which means that our sample are biased to optically-bright sources.
We are proposing to follow-up observations for WISE--SDSS photo DOGs to extend our HyLIRGs survey to more optically-faint HyLIRGs.
We will check whether or not the correlation between $i$ - [22] and $L_{\rm IR}$ can be seen for optically-faint HyLIRGs.
In that sense, this work is an important benchmark for forthcoming HyLIRGs survey. 

We are planning to summarize their spectroscopic properties and to report the relation among these quantities such as equivalent width of emission lines and infrared luminosities.
All the informations of 67 WISE--SDSS spec DOGs including such as the {\tt specobjID}, coordinates, and infrared luminosity will be presented in the forthcoming paper.

\acknowledgments
The authors gratefully acknowledge the anonymous referee for a careful reading of the manuscript and very helpful comments
We are deeply thankful to Prof. Michael A. Strauss and Dr. Masakazu A. R. Kobayashi for fruitful discussion.
This publication makes use of data products from the Wide-field Infrared Survey Explorer, which is a joint project of the University of California, Los Angeles, and the Jet Propulsion Laboratory/California Institute of Technology, funded by the National Aeronautics and Space Administration. 
Funding for SDSS-III has been provided by the Alfred P. Sloan Foundation, the Participating Institutions, the National Science Foundation, and the U.S. Department of Energy Office of Science. 
The SDSS-III web site is http://www.sdss3.org/.
SDSS-III is managed by the Astrophysical Research Consortium for the Participating Institutions of the SDSS-III Collaboration including the University of Arizona, the Brazilian Participation Group, Brookhaven National Laboratory, Carnegie Mellon University, University of Florida, the French Participation Group, the German Participation Group, Harvard University, the Instituto de Astrofisica de Canarias, the Michigan State/Notre Dame/JINA Participation Group, Johns Hopkins University, Lawrence Berkeley National Laboratory, Max Planck Institute for Astrophysics, Max Planck Institute for Extraterrestrial Physics, New Mexico State University, New York University, Ohio State University, Pennsylvania State University, University of Portsmouth, Princeton University, the Spanish Participation Group, University of Tokyo, University of Utah, Vanderbilt University, University of Virginia, University of Washington, and Yale University.
This research is also based on observations with AKARI, a JAXA project with the participation of ESA.
T. N. acknowledges financial supports from JSPS (25707010), JCG-S Scholarship Foundation, and also Institute for the Promotion of Science and Technology (Research Unit for Application of Fully Parallel Computation in Ehime University).

\end{document}